\documentclass[prd,
eqsecnum,floatfix,letterpaper,showpacs,superscriptaddress,groupedaddress,nofootinbib]{revtex4-2}

\usepackage{changes}

\def \vss{\vspace{14pt}}

\usepackage{amsmath}
\usepackage{latexsym}
\usepackage{amssymb}
\usepackage{amsfonts}
\usepackage{mathtools}
\usepackage{bm}
\usepackage{amsthm}
\usepackage{graphicx}
\usepackage{color}
\usepackage{changes}
\usepackage[normalem]{ulem}
\usepackage{natbib}
\usepackage{appendix}
\def\no{\nonumber}
\def \D{{\mathcal D}}

\def \a {\alpha}
\def \b {\beta}
\def \g {\gamma}
\def \z {\zeta}

\def \ar {\alpha_{\rm res}}
\def \br {\beta_{\rm res}}
\def \gr {\gamma_{\rm res}}

\def \Xr {X_{\rm res}}

\def \be{\begin{equation}}
\def \bea{\begin{eqnarray}}
\def \eea{\end{eqnarray}}
\def \ee{\end{equation}}

\begin{document}
\title{Second-Generation Time-Delay Interferometry}
\author{Massimo Tinto}
\email{mtinto@ucsd.edu}
\affiliation{University of California San Diego,
  Center for Astrophysics and Space Sciences,
  9500 Gilman Dr, La Jolla, CA 92093,
  U.S.A.}
\affiliation{Divis\~{a}o de Astrof\'{i}sica, Instituto
  Nacional de Pesquisas Espaciais, S. J. Campos, SP 12227-010, Brazil}
\author{Sanjeev Dhurandhar}
\email{sanjeev@iucaa.in}
\affiliation{Inter University Centre for Astronomy and Astrophysics,
  Ganeshkhind, Pune, 411 007, India}
\author{Dishari Malakar}
\email{dmkwf@umsystem.edu}
\affiliation{Missouri University of Science and Technology, Missouri S\&T, Rolla, MO 65409, U.S.A.}
\date{\today}

\begin{abstract}
  Time-Delay Interferometry (TDI) is the data processing technique
  that cancels the large laser phase fluctuations affecting the
  heterodyne Doppler measurements made by unequal-arm space-based
  gravitational wave interferometers.  The space of all TDI
  combinations was first derived under the simplifying assumption of a
  stationary array, for which the three time-delay operators
  commute. In this model, any element of the TDI space can be written
  as a linear combination of four TDI variables, the generators of the
  ``first-generation'' TDI space. To adequately suppress the laser
  phase fluctuations in a realistic array configuration, the rotation
  of the array and the time-dependence of the six inter-spacecraft
  light-travel-times have to be accounted for. In the case of the
  Laser Interferometer Space Antenna (LISA), a joint ESA-NASA mission
  characterized by slowly time varying arm-lengths, it has been
  possible to identify data combinations that, to first order in the
  inter-spacecraft velocities, either exactly cancel or suppress the
  laser phase fluctuations below the level identified by the noise
  sources intrinsic to the heterodyne measurements (the so called
  ``secondary'' noises). Here we reanalyze the problem of {\underline
    {exactly}} canceling the residual laser noise terms linear in the
  inter-spacecraft velocities. We find that the procedure for
  obtaining elements of the $2^{\rm nd}$-generation TDI space can be
  generalized in an iterative way.  This allows us to ``lift-up'' the
  generators of the $1^{\rm st}$-generation TDI space and construct
  elements of the higher order TDI space. 
\end{abstract}

\maketitle

\section{Introduction}
\label{SecI}

The Laser Interferometer Space Antenna (LISA) is a space mission
jointly proposed by the European Space Agency (ESA) and the National
Administration of Space Agency (NASA) to observe gravitational waves
(GW) in the millihertz frequency band. LISA will rely on an array of
three identical spacecraft that exchange coherent laser beams along
the three 2.5 million kilometers arms of the resulting giant (almost)
equilateral triangle.  The heliocentric trajectories of the three
spacecraft result in arm-lengths that are unequal and weakly time
dependent with inter-spacecraft relative velocities $\lesssim 10$
m/s. Since these velocities are negligible compared to the speed of
light, we are justified in retaining only first order terms in the
velocities in our considerations. The frequency noise of the LISA
stabilized lasers dominates the other secondary noises by seven or
more orders of magnitude and must be removed or sufficiently
suppressed to achieve the requisite sensitivity. By linearly combining
the appropriately delayed six one-way inter-spacecraft Doppler
measurements we can construct data combinations - the TDI combinations
- that cancel (or sufficiently suppress) the laser frequency noise
while retaining sensitivity to GWs.
\par
The simplest assumption is to regard the arm-lengths to be constant
and consider only three time-delays. This means that the light travel
time between spacecraft $i$ to $j$ is the same as between $j$ to
$i$. This is not true in general because the LISA triangle rotates
once in a year. The Sagnac effect implies that the up and down optical
paths are unequal. The TDI space that arises from the assumption of
three constant arm-lengths is the so called {\it
  $1^{\rm st}$-generation} TDI \cite{TA99,AET99,ETA00}. A rigorous
mathematical foundation for this case was laid in \cite{DNV02} proving
that the TDI space was a linear structure called in the literature as
the {\it first module of syzygies} \cite{becker,KR} which is a module
over the polynomial ring of the three time-delay operators. A neat
solution was possible because the delay operators commute and form a
commutative polynomial ring. Hilbert's theorem guarantees that in a
commutative polynomial ring over a field, all ideals are finitely
generated or the ring is Noetherian. This implies that the Gr\"obner
basis algorithm terminates and finally leads to a finite set of
generators for the module. This module is a {\it kernel} of a
homomorphism \cite{TD2020} or the TDI map the laser noise to zero and 
therefore form a null space. It has been shown that the module is generated
by a set of four generators, the simplest and most useful set being
$\a, \b, \g$ and $\z$, the Sagnac combinations.
\par
 
The next level of simplification is to consider the Sagnac effect so
that now we have six time-delays but they are considered to be
time-independent. This case can also be solved exactly \cite{TEA04,NV04} and
results in six generators for the first module of syzygies. These form
the so-called $1.5$-generation TDI space.
\par
The most general case consists of TDI combinations where the array is
rotating and the six time-delays are time dependent. In this case the
operators do not commute and one ends up with a non-commutative
polynomial ring, whose elements are strings of operators or "words" as
they are called in the literature. In the past, one of the authors
(SVD) has attempted to compute the analogous Gr\"obner basis for the
non-commutative case but found that the algorithm did not
terminate. Others have tried to use Mathematica towards the same goal
but have not succeeded. Therefore, it seems that the Gr\"obner basis
is infinite and this approach seems to be intractable. In the case of
LISA, however, the arm-lengths are slowly changing in time and the
problem therefore can be treated like a "perturbation" over the static
case and the results obtained thereby suitably generalized.
\par
In this paper, we will first study the TDI space with six different
delays that are slowly time-varying - we will consider terms only to
first order in the inter-spacecraft relative velocities. In the past
this case has been considered \cite{STEA03,TEA04,TD2020,DNV10} with
partial solutions for the so called $2^{\rm nd}$-generation TDI
space. In recent publications
\cite{muratore2020,MuratoreVetrugnoVitaleHartwig2022} an alternative
approach was proposed, in which $2^{\rm nd}$-generation TDI
combinations were obtained through the use of a computer program. Its
underlining algorithm relied on Geometric TDI \cite{Vallis} and
searched for combinations that would suppress the laser noise below
the level identified by their secondary noises. Although this approach
identified a significantly large number of $2^{\rm nd}$-generation TDI
combinations, it could not check for their independence nor assess the
dimensionality of the $2^{\rm nd}$-generation TDI space. An attempt to
answer these questions has been presented in \cite{HartwigMuratore2022},
where the new TDI channels derived in \cite{muratore2020} were related
to the Sagnac generators $\alpha, \beta, \gamma, \zeta$ of the
$1^{\rm st}$-generation TDI space.  Although the established
relationship can't be mathematically exact, it is nevertheless
accurate enough for modeling the residual noises of these
$2^{\rm nd}$-generation TDI expressions.  Its draw-back, however, is
of relying on Sagnac observables containing only the three delay
operators characteristic of a stationary array.

Finally, an analytic approach has been proposed \cite{China1} for
finding elements of the $2^{\rm nd}$-generation TDI space. It entails
a generalization of work presented in \cite{DNV10} for analytically
deriving $2^{\rm nd}$-generation unequal-arm Michelson combinations.
In \cite{China1} new Sagnac-like combinations as well as a new set of
expressions for the Monitor, Beacon, and Relay \cite{TEA04} have been
presented.

In this paper we propose instead a different approach from those cited
above for identifying TDI combinations that cancel exactly the laser
noise when the delays are characterized by small inter-spacecraft
velocities. We do so by also using only analytic techniques. Recently
matrix methods have also been employed, which lead to TDI observables
albeit numerically
\cite{PCA2006,mvallis2020,muratore2020,TDJ21}. Although the TDI
combinations we will derive in this article can be re-casted in matrix
form, we will not do that here. In our approach we first rewrite the
elements of a basis of the $1^{\rm st}$-generation TDI space in terms
of the six delay operators. Then we show that their corresponding
$2^{\rm nd}$-generation TDI expressions can be obtained by acting on
specific combinations of the data entering their expressions with
uniquely identified polynomials of the six delays. This so called
``lifting'' operation is key to our method as it allows us to
generalize the main property of a basis of the $1^{\rm st}$-generation
TDI space: elements of the $2^{\rm nd}$-generation TDI space can be
obtained by taking linear combinations of properly delayed lifted
basis. In physical terms, the operation of lifting corresponds to two
light beams each propagating clock and counterclockwise several times
around the array before being made to interfere onboard the
transmitting spacecraft. In so doing the time-variations and the
Sagnac effect on the light-travel-times get averaged out \cite{TEA04}.
\par
 
The paper is organized as follows. In section \ref{SecII} we review
some of the past relevant results, which will be required here, by
deriving a suitable set of four elements of the
$1^{\rm st}$-generation TDI space that can {\underbar{uniquely}} be
written in terms of the six time-dependent delays and generate this
space in the limit of a stationary array. Although in the stationary
configuration the basis usually adopted included the four Sagnac
combinations $\a, \b, \g, \z$, $\z$ loses its uniqueness when trying
to incorporate the six time-dependent delays in its definition. Also
it can not be interpreted as the result of the interference of two
beams that have been propagating along two different paths and a
straight forward geometric interpretation to $\zeta$ is lacking
\cite{TEA04, China1}. To avoid this complication, we use instead the
four data combinations $\a, \b, \g, X$, as generators of the
$1^{\rm st}$-generation TDI space, with $X$ being the usual
unequal-arm Michelson combination. This is possible because $\z$ is
linearly related to $\a, \b, \g, X$ \cite{AET99}.

After deriving the expressions for the residual laser noises in
specific data combinations entering the expressions of
$\a, \b, \g, X$, in section \ref{SecIII} we present useful identities
of the delay operators valid with six, time-varying, delays
characterized by small inter-spacecraft velocities. These identities
are used to derive the $2^{\rm nd}$-generation TDI combinations that
cancel the laser noise up to the velocities of the inter-spacecraft
light travel times. We call this technique ``lifting'' as it allows us
to derive the corresponding elements in the $2^{\rm nd}$-generation
TDI space by starting with the basis elements of the
$1^{\rm st}$-generation.  By then suitably delaying and linearly
combining the lifted basis of the $1^{\rm st}$-generation TDI space
one can generate elements of the higher-order space. As an application
we derive expressions of (i) $\z$-like combinations that exactly
cancel the laser noise while suppressing (like $\z$) the gravitational
wave signal in the low-part of the accessible frequency band and (ii)
$2^{\rm nd}$-generation TDIs containing only four-link measurements
(i.e. the Beacon $P_2$, Monitor $E_2$, and Relay $U_2$
combinations). In \ref{SecVI} we finally present our comments on our
findings and our conclusions.

\section{The First-Generation TDI Space}
\label{SecII}

Here we present a brief summary of the derivation of the TDI space
valid for a stationary array. We start by writing the one-way Doppler
data $y_i, y_{i'}$ in terms of the laser noises using the notation
introduced in \cite{TD2020,TDJ21}. We index the one-way Doppler data
as follows: the beam arriving at spacecraft $i$ has subscript $i$ and
is primed or unprimed depending on whether the beam is traveling
clockwise or counterclockwise around the interferometer array, with
the sense defined by a chosen orientation of the array. We define the
delay operators $\D_{i}$ by $\D_i y (t) = y (t - L_i )$ where $L_i $
is the travel-time spent by the light to travel the $i^{\rm th}$ arm
(the speed of light has been assumed to be equal to $1$). The
assumption of a stationary array implies the following expressions for
the six one-way inter-spacecraft Doppler measurements
\footnote{Besides the primary inter-spacecraft Doppler measurement
  $y_i, y_{i'}$ that contain the gravitational wave signal, other
  metrology measurements are made onboard the LISA spacecraft. This is
  because each spacecraft is equipped with two lasers and two
  proof-masses of the onboard drag-free subsystem. It has been shown
  \cite{TD2020}, however, that these onboard measurements can be
  properly delayed and linearly combined with the inter-spacecraft
  measurements to make the realistic LISA interferometry configuration
  equivalent to that of a system with only three lasers and six
  one-way inter-spacecraft measurements.}
\begin{eqnarray}
y_1 & = & \D_3  C_2 - C_1 \ \ , \ \   y_{1'} = \D_2 C_3 - C_1 \,,
\nonumber
\\
y_2 & = & \D_1 C_3 -  C_2 \ \ , \ \   y_{2'} = \D_3 C_1 - C_2 \,,
\nonumber
\\
y_3 & = & \D_2  C_1 -  C_3 \ \ , \ \   y_{3'} = \D_1 C_2 - C_3 \,.
\label{oneways}
\end{eqnarray}
The problem of identifying all possible TDI combinations associated
with the six one-way Doppler measurements becomes one of determining
six polynomials, $q_{i} , q_{i'}$, in the delay operators
$\D_i, ~i = 1, 2, 3$. The polynomials $q_i, q_{i'}$ satisfy the
following equation:
\begin{equation}
\sum_{i=1}^{3} q_i . y_i + \sum_{i'=1}^{3} q_{i'} . y_{i'} = 0 \ ,
  \label{TDI}
\end{equation}
where the above equality means ``zero laser noises''.  Equation
(\ref{TDI}) leads to the following equation for the laser noises
$C_1, C_2, C_3$:
\begin{eqnarray}
  & (& - q_1 - q_{1'} + q_3  \D_2 + q_{2'}  \D_3)  C_1
  \nonumber
  \\
  + & (& - q_2 - q_{2'} + q_1  \D_3 + q_{3'} \D_1)  C_2
        \nonumber
  \\        
  + & (& - q_3 - q_{3'} + q_2 \D_1 + q_{1'} \D_2) C_3 = 0
\label{TDIequations1}
\end{eqnarray}
Since the three random processes $C_i \ , \ i=1, 2, 3$ are
independent, the above equation can be satisfied iff the three
polynomials multiplying the three random processes are identically
equal to zero, i.e.
\begin{eqnarray}
&  & - q_1 - q_{1'} + q_3 \D_2 + q_{2'} \D_3  = 0 \ ,
  \nonumber
  \\
&  & - q_2 - q_{2'} + q_1 \D_3 + q_{3'} \D_1 = 0 \ ,
 \nonumber
  \\        
&  & - q_3 - q_{3'} + q_2 \D_1 + q_{1'} \D_2  = 0 \ .
\label{TDIequations_3}
\end{eqnarray}
The above equations apply to three time-independent arm-lengths.
\par
The resulting TDI space is the first module of syzygies obtained in
\cite{AET99,DNV02}. We will be mainly concerned with the Sagnac TDI
observables $\a, \b, \g, \z$ that generate the TDI space
\cite{AET99}.  We will also consider the Michelson TDI $X$ because of
its inherent simplicity, which will act as a guide for the other
cases. We therefore list these TDI generators below and write them as
six tuple polynomial vectors $(q_i, q_{i'})$ (in this notation the
data streams $y_i, y_{i'}$ are implicit): \bea
\a &=& (1,  \D_3, \D_3 \D_1,  -1,  - \D_2 \D_1, - \D_2) \,, \no \\
\b &=& (\D_1 \D_2, 1, \D_1, - \D_3, -1, - \D_3 \D_2) \,, \no \\
\g &=& (\D_2, \D_2 \D_3, 1, - \D_1 \D_3, - \D_1, - 1) \,, \no \\
\z &=& (\D_1, \D_2, \D_3, -\D_1, -\D_2, -\D_3) \,.
\label{generators}
\eea When the arm-lengths do not depend on time, the $\a, \b, \g, \z$
satisfy Eq. (\ref{TDIequations_3}) and therefore perfectly cancel
laser frequency noise. In this paper we propose to go beyond this
simple case, where the arm-lengths weakly depend on time. Our goal is
to generalize the $1^{\rm st}$-generation TDI space to the situation
in which the arm-lengths vary slowly.
\par

We will find that $\a$ and its cyclic permutations $\b$ and $\g$ can
be converted into second-generation TDI with the help of commutators
and some algebraic manipulation. But the TDI $\z$ is not so straight
forward as it cannot be thought of as the result of the interference
of two beams propagating along two different paths. However, we may
switch to another set of generators, namely, $\a, \b, \g$ and the
unequal-arm Michelson combination $X$. This is possible because of the
following relationship \cite{AET99} between $\z$ and $\a, \b, \g, X$:
\be \z = \D_1 X - \D_2 \D_3 \a + \D_2 \b + \D_3 \g \,,
\label{rel1}
\ee where $X$ is: \be X ~=~ (1 - \D_2^2, 0, (\D_3^2 - 1)\D_2 , \D_3^2
- 1, (1 - \D_2^2)\D_3, 0) \,.  \ee Eq. (\ref{rel1}) means that any
linear combination of the generators $\a, \b, \g, \z$ is also a linear
combination of $\a, \b, \g, X$. This implies that $\a, \b, \g$ and $X$
is another generating set for the module of syzygies. We will
therefore include the derivation of the $2^{\rm nd}$-generation
  combination $X_2$ that had already been derived in earlier
  publications \cite{TEA04,STEA03}.
\par
At the zero'th order in the inter-spacecraft velocity the laser noise
cancels out for the $1^{\rm st}$-generation TDI, those given in
Eq. (\ref{generators}) and also the Michelson $X$. But at the next
order in the velocities, the laser noise does not cancel out
completely in these TDIs, making it larger than their remaining
noises. This we call residual laser noise and denote the corresponding
TDI by the subscript {\it res}. From Eq. (\ref{TDIequations1}) we
obtain the following: \bea
\ar &=& (\D_3 \D_1 \D_2 - \D_{2} \D_{1} \D_{3}) C_1 \,, \no \\
\br &=& (\D_1 \D_2 \D_3 - \D_{3} \D_{2} \D_{1}) C_2 \,, \no \\
\gr &=& (\D_2 \D_3 \D_1 - \D_{1} \D_{3} \D_{2}) C_3 \,, \no \\
\Xr &=& (\D_3 \D_{3} \D_{2} \D_{2} - \D_{2} \D_2 \D_3 \D_{3}) C_1 \,.
\label{res}
\eea Since the above expressions contain products of operators which
are permutations of each other and occur with opposite sign, at
zero'th order the laser noise cancels out but at first order the
velocity terms (as we will see in Eq. (\ref{examples} below) multiplying the
$\dot C$ term do not cancel out. These residual laser noises must be
canceled to achieve the requisite sensitivity.

\section{TDI with six time-dependent time-delays}
\label{SecIII}

\subsection{The general model of LISA}

\begin{figure}{htbp}
\includegraphics[width = 2.5in, clip]{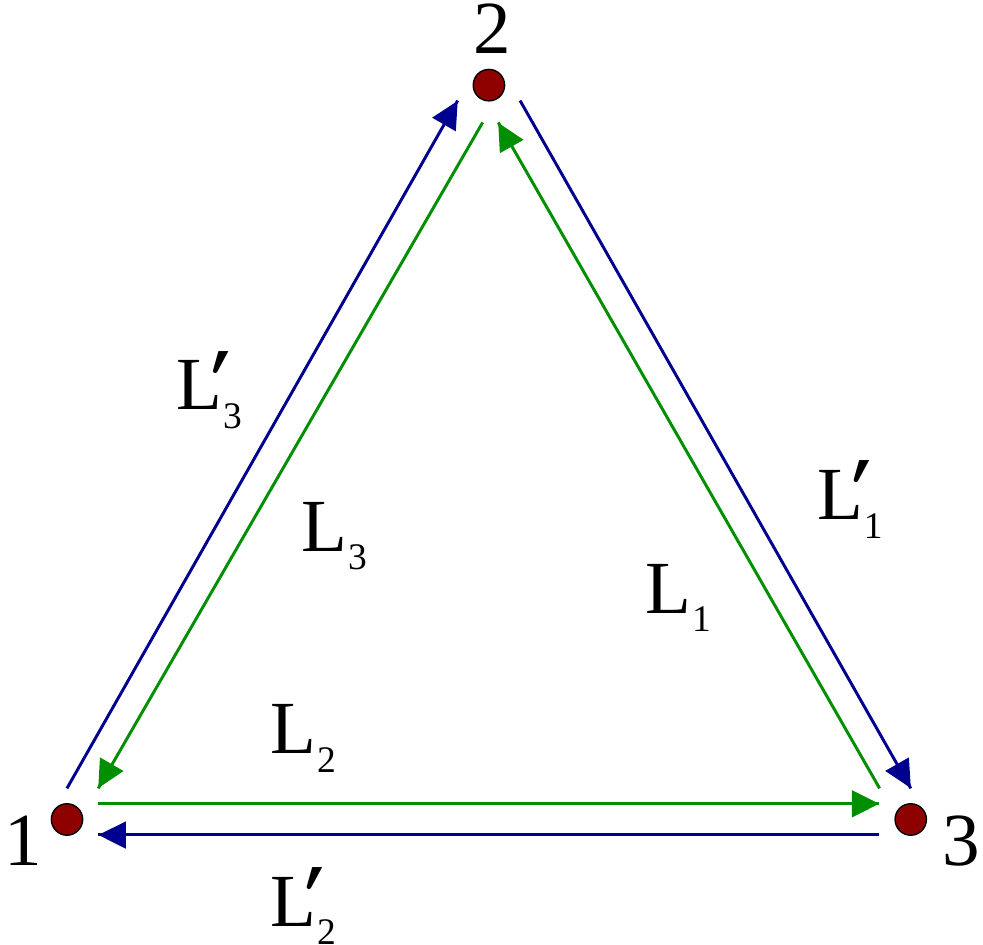}
\caption{Schematic array configuration. The spacecraft are labeled 1,
  2, and 3, and the optical paths are denoted by $L_i , L_i'$ with the
  index i corresponding to the opposite spacecraft.}
\label{fig1}
\end{figure}

We started with the $1^{\rm st}-$generation TDI because we can cleanly
derive the exact generators that completely span the TDI space. Our
idea is to use these foundational results to generalize to the
realistic model of LISA. We will achieve this by what we call as the
"lifting procedure". This procedure is described in Section
\ref{lift}. We now set up the analysis for six time-dependent
time-delays. Because of the Sagnac effect due to the rotation of the
LISA constellation, the arm-length from say spacecraft $i$ to $j$ is
not the same as the one from $j$ to $i$. Therefore $L_i \neq L'_i$ and
so we have six unequal time-dependent arm-lengths. The corresponding
operators are now labeled as $\D_i$ and $\D_{i'}$.
\par

The one-way phase measurements therefore assume the following forms
\begin{eqnarray}
y_1 & = & \D_3  C_2 - C_1 \ \ , \ \   y_{1'} = \D_{2'} C_3 - C_1 \ \ ,
\nonumber
\\
y_2 & = & \D_1 C_3 -  C_2 \ \ , \ \   y_{2'} = \D_{3'} C_1 - C_2 \ \ ,
\nonumber
\\
y_3 & = & \D_2  C_1 -  C_3 \ \ , \ \   y_{3'} = \D_{1'} C_2 - C_3 \ \ ,
\label{onewaysNew}
\end{eqnarray}
where we have adopted the labeling convention shown in
Fig. (\ref{fig1}). In it the phase difference data to be analyzed is
indexed as follows: the beam arriving at spacecraft $i$ has subscript
$i$ and is primed or unprimed depending on whether the beam is
traveling clockwise or counterclockwise (the sense defined here with
reference to Fig.~\ref{fig1}) around the array's triangle,
respectively. Thus, as seen in the figure, $y_{1}$ is the phase
difference time series measured at reception at spacecraft~1 with
transmission from spacecraft~2 (along $L_3$). The corresponding
equations analogous to Eq. (\ref{TDIequations_3}) satisfied by the
operator polynomials $q_i, q_{i'}$ generalize to:
\begin{eqnarray}
&  & - q_1 - q_{1'} + q_3 \D_2 + q_{2'} \D_{3'}  = 0 \ ,
  \nonumber
  \\
&  & - q_2 - q_{2'} + q_1 \D_3 + q_{3'} \D_{1'} = 0 \ ,
 \nonumber
  \\        
&  & - q_3 - q_{3'} + q_2 \D_1 + q_{1'} \D_{2'}  = 0 \, ,
\label{TDIequations_6}
\end{eqnarray}
where now the primed operators make an appearance.

\subsection{Slowly time-varying arm-lengths and vanishing commutators}

If the arm-lengths are time-dependent, then the operators do not
commute and the laser noise will not cancel. However, if the
arm-lengths are slowly varying we can make a Taylor expansion of the
operators and keep terms only to first order in ${\dot L_i}$ and
${\dot L'_i}$ or linear in velocities. 
\vss

Let us consider the effect of $n$ operators $\D_{k_1}, ..., \D_{k_n}$
applied on the laser noise $C(t)$. The operators could refer to either
$L_i$ or $L_{i'}$. We do not write the primes explicitly in order to
avoid clutter but the identities that we derive hold in either case.
Instead of writing $\D_{k_p}$ we may denote the same by just $k_p$
where $k_p $ take any of the values $1, 2, 3, 1', 2', 3'$. Then as
shown in \cite{DNV08,TD2020} we have:
\begin{equation}
k_n k_{n - 1} ... k_2 k_1 C(t) = C \left [t - \sum_{p = 1}^n L_{k_p}\right ]
+ \left [\sum_{j = 2}^{n} L_{k_j} \sum_{m =1}^{j-1} {\dot L_{k_m}}
                                     \right ] {\dot C} \left [t - \sum_{p = 1}^n L_{k_p} \right ]
\label{product}
\end{equation}
Let us interpret the r.h.s. of this equation. The first term is just
the laser noise at a delayed time that is equal to the sum of the
delays at time $t$. If the arm lengths were constant this would be the
only term that would be present and the operators would commute
leading to $1^{\rm st}$-generation TDI. Note that the second term {\it
  multiplies} the ${\dot C}$ evaluated at the delayed time. This term
makes the operators non-commutative. But the non-commutativity is
small because the arm-lengths are slowly varying i. e. ${\dot L} << 1$
- it is linear in the velocities. The first term is of zero'th order
in velocities. Since here we are only concerned with the second term,
we will only write the second term assuming that the zero'th order
term has been canceled exactly in the expressions. Further, in order
to avoid clutter, we will not write $C$ or ${\dot C}$ when there is no
cause for confusion. We may also write $v_{k_p} = {\dot
  L_{k_p}}$. Then with this understanding we may write
Eq. (\ref{product}) as: \be k_n k_{n - 1} ... k_2 k_1 = \sum_{j =
  2}^{n} L_{k_j} \sum_{m = 1}^{j -1} v_{k_m} \,.
\label{mastereq}
\ee Note that the $k_p$ need not be distinct - the operators may
repeat. To write down the first few products explicitly: \bea
\D_2 \D_1 &=& L_2 v_1 \,, \no \\
\D_3 \D_2 \D_1 &=& L_2 v_1 + L_3 (v_1 + v_2) \,, \no \\
\D_4 \D_3 \D_2 \D_1 &=& L_2 v_1 + L_3 (v_1 + v_2) + L_4 (v_1 + v_2 +
v_3) \,.
\label{examples}
\eea It was shown in \cite{DNV10,TD2020} that certain commutators
cancel the laser noise under the approximation we are making. Let
$x_1, x_2, ..., x_n$ and $z_1, z_2, ..., z_n$ be delay operators. Then
it follows from Eq. (\ref{mastereq}) that, \be [x_1 x_2 ...x_n, z_1
z_2 ...z_n] = \sum_{k = 1}^n L_{x_k} \sum_{m=1}^n v_{z_m} - \sum_{m =
  1}^n L_{x_m} \sum_{k=1}^n v_{z_k} \,.
\label{commutator}
\ee Let $\sigma$ be a permutation on $n$ symbols. Then
$x_{\sigma(1)}, x_{\sigma(2)}, ..., x_{\sigma(n)}$ is a permutation of
$x_1, x_2, ..., x_n$, then it is easy to show that, \be [x_1 x_2
...x_n, x_{\sigma(1)}, x_{\sigma(2)}, ..., x_{\sigma(n)}] = 0 \,.
\label{commutator2}
\ee It was shown in \cite{DNV10} that a large number of Michelson type
TDI can be generated by relying on Eq. (\ref{commutator2}) and, more recently
\cite{China1}, those results have been generalized to find many other
elements of the $2^{\rm nd}$-generation TDI space such as the Sagnac,
Symmetric Sagnac, Monitor, Beacon and Relay.
\par

\section{The lifting procedure}
\label{lift}

We first need to derive the expressions of the four generators,
$\a, \b, \g, X$, of the $1^{\rm st}$-generation TDI formulation that
include the six delays $i, i' \ \ i, i'= 1, 2, 3, 1', 2', 3'$. Since
these combinations correspond to beams propagating clock and
counterclockwise, we can then generalize the procedure for identifying
combinations that suppress the laser noise to the required levels
\cite{STEA03,TEA04}. This is done by making each beam propagate clock
and counterclockwise a number of times such that the resulting data
combinations exactly cancel the laser noise up to the velocities of
the six delays. This procedure, which we now call ``lifting'', is
unique and can be applied iteratively an arbitrary number of times.
It should be emphasized that some elements of the
$2^{\rm nd}$-generation TDI space, like the Sagnac combinations
$\a, \b, \g$, require more than two ``lifting'' iterations to exactly
cancel the laser noise up to the velocity terms
\cite{TEA04,TD2020}. Therefore we will refer to the space of the
$2^{\rm nd}$-generation TDI space as those combinations that
{\underline {exactly}} cancel the laser noise up to the velocities
terms.

\subsection{The Unequal-arm Michelson $X$}

The $X$ combination includes the four one-way Doppler measurements,
($y_1, y_{1'}, y_{2'}, y_{3}$) from the two arms centered on
spacecraft 1.  In what follows we will present the method discussed in
\cite{STEA03,TEA04,TD2020} for obtaining the $2^{\rm nd}$-generation
TDI $X_2$, and generalize this approach to derive other unequal-arm
Michelson combinations.  Let us consider the following synthesized
two-way Doppler measurements

\begin{align}
X_{\uparrow} &\equiv y_1 + \D_3 y_{2'} = (\D_{3} \D_{3'} - I) \ C_1 \ ,
\nonumber
\\
X_{\downarrow} &\equiv y_{1'} + \D_{2'} y_3 = (\D_{2'} \D_{2} - I) \
                 C_1 \ .
\end{align}
As we know, the $1^{\rm st}$-generation TDI combination $X$, is
equal to the following expression
\begin{equation}
X \equiv (\D_{3} \D_{3'} - I) \ X_{\downarrow} - (\D_{2'} \D_{2} - I) \ X_{\uparrow}
= [\D_{3} \D_{3'} , \D_{2'} \D_{2}] \ C_1 \ .
\label{X}
\end{equation}
It is easy to see the above commutator is different from zero when the
delays are functions of time and, to first order, is in fact
  proportional to the inter-spacecraft relative velocities. To derive
the $2^{\rm nd}$-generation TDI combination $X_2$, which cancels
exactly the laser noise up to velocity terms, we rewrite the above
expression for $X$ in terms of its two synthesized beams. They are
equal to
\begin{align}
X_{\uparrow \uparrow} &\equiv  \D_{2'} \D_{2} \ X_{\uparrow} + \
                        X_{\downarrow} = (\D_{2'} \D_{2} \D_{3} \D_{3'} - I) \ C_1 \ ,
\nonumber
\\
X_{\downarrow \downarrow} &\equiv  \D_{3} \D_{3'} \ X_{\downarrow}  +  X_{\uparrow} = 
(\D_{3} \D_{3'} \D_{2'} \D_{2} - I) \ C_1 \ ,
\end{align}
The $X_2$ expression can be derived by repeating the same
procedure used for deriving $X$. This results in the following
expression
\begin{equation}
X_2 \equiv (\D_{3} \D_{3'} \D_{2'} \D_{2} - I) X_{\uparrow \uparrow} -
(\D_{2'} \D_{2} \D_{3} \D_{3'} - I) X_{\downarrow \downarrow} =
[\D_{3} \D_{3'} \D_{2'} \D_{2}, \D_{2'} \D_{2} \D_{3} \D_{3'}] \ C_1 =
0 \ ,
\label{X2}
\end{equation}
where the equality to zero means ``up to velocity terms'', and it is
consequence of the general property of the commutators of the delay
operators proved in the previous section. It is clear that the
iterative procedure we have implemented for deriving both $X$ and
$X_2$ can be repeated to obtain the expression for other unequal-arm
Michelson combinations. Lastly we note that, because the magnitudes of
the frequency fluctuations associated with a GW signal and the
secondary noises in $X_2$ are significantly smaller than those of a
laser, the commutator of two delay operators applied to them results
in relative frequency fluctuations that are about seven orders of
magnitude smaller than their values and can therefore be regarded as
equal to zero. This means that the order by which two delay operators
act on a GW signal and the secondary noises can be ignored. This
observation implies that their contributions to $X_2$,
$X_2^{\rm GW,N}$, are related to those in $X$, $X^{GW,N}$, through the
following relationship
\begin{equation}
X_2^{\rm GW,N} = (I - \D_3 \D_{3'} \D_{2'}\D_2) \ X^{\rm GW,N} \ .
\label{X2GWN}
\end{equation}
Eq. (\ref{X2GWN}) follows from Eqs. (\ref{X2}, \ref{X}) after some
simple algebraic manipulations that account for the commutativity of
the delay operators when applied to a GW signal and the secondary
noises. It states the GW signal and secondary noises present in $X_2$
are related to those in $X$ through the operator
$(I - \D_3 \D_{3'} \D_{2'}\D_2)$.  It also says that the GW
sensitivity of $X_2$ is equal to that of $X$ because the Fourier
transfer function of the operator $(I - \D_3 \D_{3'} \D_{2'}\D_2)$
multiplies both the GW signal and the noise in $X$ and thus cancels
out. In general, if $A$ and $B$ are two TDI observables such that
$A = p (\D_i, \D_{i'}) B$, where $p (\D_i, \D_{i'})$ is a polynomial
in the delay operators $\D_i$ and $\D_{i'}$ then because the same
transfer function scales both the signal and the noise in $A$ and $B$,
the sensitivities of $A$ and $B$ are identical.

We will be using Eq. (\ref{X2GWN}) later on when deriving other
$2^{\rm nd}$-generation TDI combinations.
\par

\subsection{The Sagnac combination $\alpha$}

The $\alpha$ combination represents a synthesized Sagnac
interferometer. In it two synthesized light-beams interfere onboard
spacecraft $1$ after making a clock and counterclockwise loop around
the array.  From simple geometric considerations on the delays and
paths traveled by the two synthesized beams it is easy to derive the
following expression for $\alpha$
\begin{equation}
\alpha = [y_1 + \D_3 y_2 + \D_3 \D_1 y_3] - [y_{1'} + \D_{2'} y_{3'} +
\D_{2'} \D_{1'} y_{2'}] \ .
\label{alpha}
\end{equation}
After substituting Eqs.(\ref{onewaysNew}) into Eq. (\ref{alpha}) we
find the expression of the residual laser noise $C_1(t)$ in $\alpha$
to be equal to
\begin{equation}
  \alpha_{res} = (\D_3 \D_1 \D_2 - \D_{2'} \D_{1'} \D_{3'}) C_1
  \label{alphares}
\end{equation}
The $1.5$-generation TDI Sagnac observables were obtained by making
each beam go around the array twice, in the clock and counterclockwise
directions. In so doing the effects of rotation could get canceled
exactly, while linear terms in the velocities multiplying the laser
noise would get adequately suppressed below the secondary noises. As
we will show below, exact cancellation of the laser noise up to
velocity terms can be achieved by having the beams make additional
loops around the array.

Let us consider the two beams forming alpha
\begin{align}
\alpha_{\uparrow}  & \equiv y_1 + D_3 y_2 + D_3 D_1 y_3  = (D_3 D_1 D_2 - I)
           C_1 \ ,
  \nonumber
\\
\alpha_{\downarrow} &\equiv y_{1'} + D_{2'} y_{3'} + D_{2'} D_{1'} y_{2'} =
           (D_{2'} D_{1'} D_{3'} - I) C_1 \ ,
\label{rho}
\end{align}
The $1.5$-generation TDI Sagnac observable, $\alpha_{1.5}$, was then
obtained by forming the following linear combination of
$\alpha_{\uparrow}$ and $\alpha_{\downarrow}$
\begin{equation}
  \alpha_{1.5} \equiv (D_{2'} D_{1'} D_{3'} - I) \alpha_{\uparrow}  - (D_3 D_1 D_2 -  I) \alpha_{\downarrow}
  = [D_{2'} D_{1'} D_{3'}, D_3 D_1 D_2] C_1 \ .
  \label{alphaalpha}  
\end{equation}
From the properties of commutators derived in the previous section, we
recognize that the right-had-side of Eq. (\ref{alphaalpha}) does not
cancel the laser noise terms containing the velocities
\footnote{Although the 1.5-generation $\alpha$ combination was also
  referred to in the literature as being an element of the
  $2^{\rm nd}$-generation TDI space because it suppresses the laser
  noise below the secondary noises, here we will refer to it as
  $\alpha_{1.5}$ since it does not exactly cancel the laser noise up
  to the velocity terms.}.  However, by applying our iterative
procedure one more time this can be achieved. Let us first write the
following two expressions, which take into account
Eq. (\ref{alphaalpha})
\begin{eqnarray}
\alpha_{\uparrow \uparrow} & = & D_{2'} D_{1'} D_{3'} \  \alpha_{\uparrow} + \alpha_{\downarrow} = (D_{2'} D_{1'} D_{3'} D_3 D_1 D_2 - I) C_1 \ ,
\nonumber
  \\
\alpha_{\downarrow \downarrow} & = & \alpha_{\uparrow} + D_3 D_1 D_2 \ \alpha_{\downarrow} = (D_3 D_1 D_2 D_{2'} D_{1'} D_{3'} - I) C_1 \ .
\label{rhorho}
\end{eqnarray}
From the above equation we then obtain the following expression for $\alpha_2$
\begin{eqnarray}
\alpha_2 &\equiv& (D_3 D_1 D_2 D_{2'} D_{1'} D_{3'} - I) \alpha_{\uparrow \uparrow}
                  - (D_{2'} D_{1'} D_{3'} D_3 D_1 D_2 - I)
                  \alpha_{\downarrow \downarrow} 
             \nonumber
             \\
             &=& [D_{2'} D_{1'} D_{3'} D_3 D_1 D_2, D_3 D_1 D_2 D_{2'} D_{1'} D_{3'}] C_1 \ .
\label{alpha2}
\end{eqnarray}
We may notice that the operator that applies to $C_1$ in
Eq. (\ref{alpha2}) is the commutator of two delay operators, each
being the product of the same number of primed and unprimed delay
operators and related by permutations of their indices. From the
commutator identities derived in the previous section, we conclude
that such a commutator results in the exact cancellation of the laser
noise up to velocity terms.  The iterative process highlighted above
can of course be repeated, resulting in other TDI combinations
Finally we now provide the expression of $\alpha_2^{\rm GW, N}$ in
terms of $\alpha^{\rm GW,N}$, which follows from Eqs. (\ref{alpha2},
\ref{rhorho}, \ref{alphaalpha}, \ref{rho})
\begin{equation}
  \alpha_2^{\rm GW, N} = (\D_3 \D_1 \D_2 \D_{2'} \D_{1'} \D_{3'} - I) \
  (\D_{2'} \D_{1'} \D_{3'} - I) \ \alpha^{\rm GW,N} \ .
  \label{alpha2GWN}
\end{equation}
Eq. (\ref{alpha2GWN}) above reflects the fact that the delay operators
can be treated as constant and that the inequality between the primed
and unprimed delays can also be disregarded when acting on a GW signal
and the secondary noises in $\alpha_2$. Like in the case of $X_2$ and
$X$, here too $\a_2$ and $\a$ have the same sensitivity to
gravitational waves as the same Fourier transfer function multiplies
both the GW signal and the secondary noises of $\a$.

As it will be shown below, Eqs. (\ref{alpha2GWN}, \ref{X2GWN}) will
play a key role in the derivation of other $2^{\rm nd}$-generation TDI
combinations by properly delaying and linearly combining the four
observables ($\alpha_2, \beta_2, \gamma_2, X_2$).
\section{The $2^{\rm nd}$-generation TDI space}

In what follows we derive the expressions for the symmetric Sagnac
combination $\z_2$, the Monitor $E_2$, the Beacon $P_2$, and the Relay
$U_2$ by taking specific combinations of the lifted basis
($\alpha_2, \beta_2, \gamma_2, X_2$). Although there already exist
expressions in the literature for the $E_2$, $P_2$, $U_2$ combinations
that cancel the laser noise up to velocity terms \cite{TEA04,TD2020},
the $\z_{1.5}$ combinations only suppresses the laser noise below the
secondary noises. Also, our derivations below will show that their
expressions are not unique and that there exist in fact an infinite
number of them displaying sensitivities to GWs equal to their
corresponding $1^{\rm st}$-generation counterpart.

\subsection{The Sagnac combination $\z$}

With the expressions of ($\alpha_2, \beta_2, \gamma_2, X_2$) derived
in the previous sections, we can now generate other elements of the
$2^{\rm nd}$-generation TDI space by taking linear combinations of
($\alpha_2, \beta_2, \gamma_2, X_2$) with coefficients being
polynomials in the six delay operators. Since the four polynomials can
be arbitrary, we conclude that we can generate an $\infty^4$ number of
elements of the $2^{\rm nd}$-generation TDI space.

In the case of almost equilateral arrays (with LISA being the most well 
known example) among all TDI combinations the symmetric Sagnac $\z$ is
characterized by a GW transfer function that suppresses the GW signal
in the lower part of the accessible frequency band. By being still
affected by the instrumental noise sources, $\z$ has been shown to
provide future space-based GW interferometers with the capability of
calibrating their in-flight noise performance in the presence of a
strong astrophysical GW background \cite{TAE01}.

Expressions for $\z$ that could exactly cancel the laser noise in the
case of a rigidly rotating array were found in the literature
\cite{TEA04,NV04}. They have also been shown to adequately suppress
the laser noise below the secondary noise sources in the case of
slowing varying arm-lengths. Here we show that it is possible to
derive a family of $\z$-like combinations that exactly cancel the
laser noise up to velocity terms by taking specific linear combinations
of ($\alpha_2, \beta_2, \gamma_2, X_2$).  Let us first write the
following general linear combination of
($\alpha_2, \beta_2, \gamma_2, X_2$)
\begin{equation}
  \zeta_2 \equiv \lambda_X X_2 + \lambda_\alpha \alpha_2 + \lambda_\beta \beta_2 +
  \lambda_\gamma \gamma_2 \ ,
  \label{zeta2}
\end{equation}
where the four polynomials of the delay operators, ($\lambda_X, \lambda_\alpha,
\lambda_\beta, \lambda_\gamma$) are at the moment unknown.

Since ($\alpha_2, \beta_2, \gamma_2, X_2$) cancel exactly the laser
noises, it is clear that any linear combination of them (such as that
given in Eq. (\ref{zeta2})) is also laser noise-free.  Since
($\alpha_2, \beta_2, \gamma_2, X_2$) now only contain the GW signal and
the secondary noises, we can replace in Eq. (\ref{zeta2}) their
expressions in terms of the $1^{\rm st}$-generation TDI combinations
as given by Eqs.(\ref{X2GWN}, \ref{alpha2GWN}). This results in the
following expression for $\zeta_2^{\rm GW,N}$
\begin{equation}
  \zeta_2^{\rm GW,N} = \lambda_X (I - \D_3 \D_{3'} \D_{2'} \D_2) X^{\rm GW,N} +
  (\D_3 \D_1 \D_{2} \D_{2'} \D_{1'} \D_{3'} - I) (\D_{2'} \D_{1'}
  \D_{3'} - I) [\lambda_\alpha \alpha^{\rm GW,N} + \lambda_\beta \beta^{\rm GW,N} + \lambda_\gamma \gamma^{\rm GW,N}] \ .
  \label{zeta22}
\end{equation}
Since
$\zeta^{\rm GW,N} = \D_1 X^{\rm GW,N} - \D_2 \D_3 \alpha^{\rm GW,N} +
\D_2 \beta^{\rm GW,N} + D_3 \gamma^{\rm GW,N}$, it is then easy to
identify the following expressions for the polynomials
($\lambda_X, \lambda_\alpha, \lambda_\beta, \lambda_\gamma$) that guarantee $\zeta_2$ to have
the same sensitivity as $\zeta$
\begin{eqnarray}
  \lambda_X & = &   (\D_3 \D_1 \D_{2} \D_{2'} \D_{1'} \D_{3'} - I) (\D_{2'} \D_{1'}
            \D_{3'} - I) \D_1 \ ,
            \nonumber
  \\
  \lambda_\alpha & = & - (I - \D_3 \D_{3'} \D_{2'} \D_2) \D_2 \D_3 \ ,
                 \nonumber
  \\
    \lambda_\beta & = & (I - \D_3 \D_{3'} \D_{2'} \D_2) \D_2 \ ,
                 \nonumber
  \\
    \lambda_\gamma & = &  (I - \D_3 \D_{3'} \D_{2'} \D_2) \D_3 \ .
                   \label{PolLam}
\end{eqnarray}
Note the above four polynomials are not unique as they are defined up
to an arbitrary polynomial multiplying them. If we now take the above
expressions for ($\lambda_X, \lambda_\alpha, \lambda_\beta,
\lambda_\gamma$) and substitute them into Eq. (\ref{zeta2}) we obtain
the final expressions for $\z_2$ and $\z_2^{GW,N}$
\begin{equation}
  \zeta_2 = (\D_3 \D_1 \D_{2} \D_{2'} \D_{1'} \D_{3'} - I) (\D_{2'} \D_{1'}
  \D_{3'} - I) \D_1  X_2 + (I - \D_3 \D_{3'} \D_{2'} \D_2) [- \D_2 \D_3 \alpha_2
+  \D_2 \beta_2 +  \D_3 \gamma_2 ] \ ,
  \label{zeta222}
\end{equation}
\begin{equation}
\z_2^{GW,N} = (\D_3 \D_1 \D_{2} \D_{2'} \D_{1'} \D_{3'} - I) (\D_{2'} \D_{1'}
\D_{3'} - I) (I - \D_3 \D_{3'} \D_{2'} \D_2) \z^{GW,N}
\label{zeta222GWN}
\end{equation}
As expected from the criterion adopted for identifying the four
polynomials
($\lambda_X, \lambda_\alpha, \lambda_\beta, \lambda_\gamma$),
Eq. (\ref{zeta222GWN}) explicitly shows that $\z_2$ has the same
sensitivity to GWs as $\z$. This is because the Fourier components of
the GW signals and the secondary noises in $\z_2$ have the same
transfer function to the GW signal and the secondary noises in $\z$.

\subsection{The Monitor $E_2$  Combinations}

The monitor is a TDI combination that relies on only four Doppler
measurements. As the name suggests, it corresponds to an array
configuration in which one spacecraft can only receive laser light
from the other two.  To derive the $2^{\rm nd}$-generation TDI
expression for such configuration, we first remind the reader that the
$1^{\rm st}$-generation TDI combination $E$ is related to the basis
elements ($\a, \b, \g, X$) through the following relationship
\cite{ETA00}
\begin{equation}
  E = \alpha - \D_1 \z = \alpha - \D_1 (\D_1 X - \D_2 \D_3 \alpha +
  \D_2 \beta + \D_3 \gamma) = - \D_1 \D_1 X  + (I + \D_1 \D_2 \D_3) \alpha 
-  \D_1 \D_2 \beta - \D_1 \D_3 \gamma \ ,
  \label{E}
\end{equation}
where we have substituted the expression for $\z$ in terms of ($\a,
\b, \g, X$) given in Eq.(\ref{rel1}).

Let us now write again the following general linear combination of
($\alpha_2, \beta_2, \gamma_2, X_2$)
\begin{equation}
  E_2 \equiv \mu_X X_2 + \mu_\alpha \alpha_2 + \mu_\beta \beta_2 +
  \mu_\gamma \gamma_2 \ ,
  \label{E2}
\end{equation}
where the four polynomials of the delay operators,
($\mu_X, \mu_\alpha, \mu_\beta, \mu_\gamma$), are unknown.

Since ($\alpha_2, \beta_2, \gamma_2, X_2$) cancel exactly the laser
noises, any linear combination of them (such as that
given in Eq. (\ref{E2})) is also laser noise-free.  Since
($\alpha_2, \beta_2, \gamma_2, X_2$) now only contain the GW signal and
the secondary noises, we can replace in Eq. (\ref{E2}) their
expressions in terms of the $1^{\rm st}$-generation TDI combinations
as given by Eqs.(\ref{X2GWN}, \ref{alpha2GWN}). This results in the
following expression for $E_2^{\rm GW,N}$
\begin{equation}
E_2^{\rm GW,N} = \mu_X (I - \D_3 \D_{3'} \D_{2'} \D_2) X^{\rm GW,N} +
  (\D_3 \D_1 \D_{2} \D_{2'} \D_{1'} \D_{3'} - I) (\D_{2'} \D_{1'}
  \D_{3'} - I) [\mu_\alpha \alpha^{\rm GW,N} + \mu_\beta \beta^{\rm GW,N} + \mu_\gamma \gamma^{\rm GW,N}] \ .
  \label{E22}
\end{equation}
Since
$E^{\rm GW,N} = - \D_1 \D_1 X^{\rm GW,N} + (I + \D_1 \D_2 \D_3)
\alpha^{\rm GW,N} - \D_1 \D_2 \beta^{\rm GW,N} - \D_1 \D_3 \gamma^{\rm
  GW,N} $, it is then easy to derive the following expressions for
($\mu_X, \mu_\alpha, \mu_\beta, \mu_\gamma$) that guarantee $E_2$ to have
the same sensitivity as $E$
\begin{eqnarray}
  \mu_X & = &  - (\D_3 \D_1 \D_{2} \D_{2'} \D_{1'} \D_{3'} - I) (\D_{2'} \D_{1'}
            \D_{3'} - I) \D_1 \D_1 \ ,
            \nonumber
  \\
  \mu_\alpha & = & (I - \D_3 \D_{3'} \D_{2'} \D_2) (I + \D_1 \D_2 \D_3)  \ ,
                 \nonumber
  \\
    \mu_\beta & = & - (I - \D_3 \D_{3'} \D_{2'} \D_2) \D_1 \D_2 \ ,
                 \nonumber
  \\
    \mu_\gamma & = &  - (I - \D_3 \D_{3'} \D_{2'} \D_2) \D_1 \D_3 \ ,
\label{PolMu}
\end{eqnarray}

As in the case of $\z_2$, the four polynomials identifying $E_2$ are
also not unique as they are defined up to an arbitrary polynomial
multiplying them. In other words, there exist an infinite number
of Monitor combinations in the $2^{\rm nd}$-generation TDI space.

If we now substitute the expressions above for the polynomials 
($\mu_X, \mu_\alpha, \mu_\beta, \mu_\gamma$)  into Eq. (\ref{E2}), we
obtain the following expressions for $E_2$ and $E_2^{GW,N}$
\begin{equation}
  E_2 = - (\D_3 \D_1 \D_{2} \D_{2'} \D_{1'} \D_{3'} - I) (\D_{2'} \D_{1'}
  \D_{3'} - I) \D_1 \D_1 X_2
  + (I - \D_3 \D_{3'} \D_{2'} \D_2) [(I + \D_1 \D_2 \D_3)   \alpha_2
  - \D_1 \D_2  \beta_2
  - \D_1 \D_3   \gamma_2] \ ,
  \label{E222}
\end{equation}
\begin{equation}
  E_2^{GW,N} = (\D_3 \D_1 \D_{2} \D_{2'} \D_{1'} \D_{3'} - I) (\D_{2'}
  \D_{1'}   \D_{3'} - I) (I - \D_3 \D_{3'} \D_{2'} \D_2) E^{GW,N}
\end{equation}

\subsection{The Beacon $P_2$  Combinations}
The beacon, like the monitor, is a TDI combination that relies on only
four Doppler measurements. As the name suggests, it corresponds to an
array configuration in which one spacecraft can only transmit
laser light to the other two but is unable to receive from them.  As
in the case of the monitor combination, we first observe that the
$1^{\rm st}$-generation TDI combination $P$ is related to the basis
elements ($\a, \b, \g, X$) through the following relationship \cite{ETA00}
\begin{equation}
  P = \z - \D_1 \a = \D_1 X - \D_2 \D_3 \alpha +  \D_2 \beta + \D_3
  \gamma - \D_1 \a = \D_1 X - (D_1 + \D_2 \D_3) \alpha +  \D_2 \beta + \D_3
  \gamma \ ,
  \label{P}
\end{equation}
where again we have taken advantage of the expression for $\z$ in
terms of ($\a, \b, \g, X$) given in Eq.(\ref{rel1}).

As done in the previous subsection, we first take a linear combination
of ($\alpha_2, \beta_2, \gamma_2, X_2$) with four unknown polynomials
($\nu_X, \nu_\alpha, \nu_\beta, \nu_\gamma$) of the delay operators
\begin{equation}
  P_2 \equiv \nu_X X_2 + \nu_\alpha \alpha_2 + \nu_\beta \beta_2 +
  \nu_\gamma \gamma_2 \ .
  \label{P2}
\end{equation}

Since ($\alpha_2, \beta_2, \gamma_2, X_2$) cancel exactly the laser
noises, any linear combination of them (such as that given by
Eq. (\ref{P2})) is also laser noise-free. This implies that 
we can replace in Eq. (\ref{P2}) their
expressions in terms of the $1^{\rm st}$-generation TDI combinations
as given by Eqs.(\ref{X2GWN}, \ref{alpha2GWN}). This results in the
following expression for $P_2^{\rm GW,N}$
\begin{equation}
P_2^{\rm GW,N} = \nu_X (I - \D_3 \D_{3'} \D_{2'} \D_2) X^{\rm GW,N} +
  (\D_3 \D_1 \D_{2} \D_{2'} \D_{1'} \D_{3'} - I) (\D_{2'} \D_{1'}
  \D_{3'} - I) [\nu_\alpha \alpha^{\rm GW,N} + \nu_\beta \beta^{\rm GW,N} + \nu_\gamma \gamma^{\rm GW,N}] \ .
  \label{P22}
\end{equation}
Since $P^{\rm GW,N} = \D_1 X^{\rm GW,N} - (\D_1 + \D_2 \D_3)
\alpha^{\rm GW,N} + \D_2 \beta^{\rm GW,N} + \D_3 \gamma^{\rm
  GW,N} $, it is then easy to recognize the following expressions for
($\nu_X, \nu_\alpha, \nu_\beta, \nu_\gamma$) guarantee
$P_2$ to have the same sensitivity as $P$
\begin{eqnarray}
  \nu_X & = &   (\D_3 \D_1 \D_{2} \D_{2'} \D_{1'} \D_{3'} - I) (\D_{2'} \D_{1'}
            \D_{3'} - I) \D_1 \ ,
            \nonumber
  \\
  \nu_\alpha & = & - (I - \D_3 \D_{3'} \D_{2'} \D_2)  (\D_1 + \D_2 \D_3) \ ,
                 \nonumber
  \\
    \nu_\beta & = & (I - \D_3 \D_{3'} \D_{2'} \D_2) \D_2 \ ,
                 \nonumber
  \\
    \nu_\gamma & = &  (I - \D_3 \D_{3'} \D_{2'} \D_2) \D_3 \ .
\label{PolNu}
\end{eqnarray}

The above four polynomials are again defined up to an arbitrary
polynomial multiplying them. By substituting them into Eq. (\ref{P2})
we finally  get
\begin{equation}
P_2 =  (\D_3 \D_1 \D_{2} \D_{2'} \D_{1'} \D_{3'} - I) (\D_{2'} \D_{1'}
\D_{3'} - I) \D_1 X_2
+ (I - \D_3 \D_{3'} \D_{2'} \D_2)  [-(D_1 + \D_2 \D_3) \a_2 + \D_2
\b_2 + \D_3 \g_2] \ ,
\label{P222}
\end{equation}
and
\begin{equation}
P_2^{GW,N} =  (\D_3 \D_1 \D_{2} \D_{2'} \D_{1'} \D_{3'} - I) (\D_{2'}
\D_{1'} \D_{3'} - I) (I - \D_3 \D_{3'} \D_{2'} \D_2) P^{GW,N}
\end{equation}

\subsection{The Relay $U_2$  Combinations}
The relay is a TDI combination corresponding to an array configuration
in which one spacecraft can only receive along one arm and
transmit along the other. As in the case of the previous two four-link
combinations, we first observe that the $1^{\rm st}$-generation TDI
combination $U$ is related to the basis elements ($\a, \b, \g, X$)
through the following relationship \cite{ETA00}
\begin{equation}
  U = \D_1 \g - \b  \ .
  \label{U}
\end{equation}
Given the above form of $U$, the most general expression for $U_2$
will be determined by the following linear combination of $\b_2$ and
$\g_2$
\begin{equation}
  U_2 \equiv \delta_\beta \beta_2 +  \delta_\gamma \gamma_2 \ ,
  \label{U2}
\end{equation}
where $\delta_\b , \delta_\g$ are unknown polynomials of the delay
operators. Since ($\beta_2, \gamma_2$) cancel exactly the laser
noises, any linear combination of them (such as that given by
Eq. (\ref{U2})) is also laser noise-free. This implies that we can
replace in Eq. (\ref{U2}) their expressions in terms of the
$1^{\rm st}$-generation TDI combinations as given by
Eq. (\ref{alpha2GWN}). This results in the following expression for
$U_2^{\rm GW,N}$
\begin{equation}
U_2^{\rm GW,N} =  (\D_3 \D_1 \D_{2} \D_{2'} \D_{1'} \D_{3'} - I) (\D_{2'} \D_{1'}
  \D_{3'} - I) [\delta_\beta \beta^{\rm GW,N} + \delta_\gamma \gamma^{\rm GW,N}] \ .
  \label{U22}
\end{equation}
Since $U^{\rm GW,N} = - \beta^{\rm GW,N} + \D_1 \gamma^{\rm GW,N} $,
it is easy to identify the following expressions for
($\delta_\beta , \delta_\gamma$) guarantee $U_2$ to have the same
sensitivity of $U$
\begin{eqnarray}
    \delta_\beta & = & - 1 \ ,
                 \nonumber
  \\
    \delta_\gamma & = &  \D_1 \ .
\label{PolDelta}
\end{eqnarray}
The above two polynomials are defined up to an arbitrary polynomial
multiplying them.  The resulting expressions for $U_2$ and $U_2^{GW,N}$
are therefore equal to
\begin{equation}
  U_2 = -\b_2 + D_1 \g_2 \ ,
  \label{U222}
\end{equation}
and
\begin{equation}
U_2^{GW,N} =  (\D_3 \D_1 \D_{2} \D_{2'} \D_{1'} \D_{3'} - I) (\D_{2'} \D_{1'} \D_{3'} - I) U^{GW,N}
\end{equation}
  
\section{Conclusions}
\label{SecVI}

We revisited the $2^{\rm nd}$-generation TDI space, i.e. the set of
TDI combinations canceling the laser noise up to terms linear in the
time-derivatives of the inter-spacecraft light-travel-times. We
identified analytic expressions for the Sagnac ($\a_2, \b_2, \g_2$)
and unequal-arm Michelson combination $X_2$ that exactly cancel the
laser noises up to linear terms in the inter-spacecraft
velocities. Our derivation relies on an iterative procedure we named
``lifting''. This technique entails making two synthesized laser beams
go around the array along clock and counterclockwise paths a number of
times before interfering back at the transmitting spacecraft. We found
that, to cancel the laser phase fluctuations (up to velocity terms) in
the Sagnac combination $\a$, the two synthesized beams need to make at
least three loops around the array before interfering back at the
transmitting spacecraft. By relying on the expressions of the lifted
Sagnac, ($\a_2, \b_2, \g_2$), and unequal-arm Michelson combinations,
$X_2$, we were able to identify an infinite number of expressions for
$\z$-like, Monitor, Beacon, and Relay combinations. This was done by
taking linear combinations of ($\a_2, \b_2, \g_2, X_2$) with
polynomials of the delay operators that result in TDI combinations
whose sensitivities equal those of their $1^{\rm st}$-generation
counterparts. In this regard we can say of having identified a mapping
between the $1^{\rm st}$- and the $2^{\rm nd}$-generation TDI spaces
by which any element of the $1^{\rm st}$-generation TDI space is
lifted up. 
\par

We believe the iterative procedure so effectively employed in this
article may be extended to cancel the laser frequency noise at higher
orders. We will follow up on these ideas in our forthcoming
investigations.

\section*{Acknowledgments}

M.T. thanks the Center for Astrophysics and Space Sciences (CASS) at
the University of California San Diego (UCSD, U.S.A.) and the National
Institute for Space Research (INPE, Brazil) for their kind hospitality
while this work was done. S.V.D. acknowledges the support of the
Senior Scientist Platinum Jubilee Fellowship from National Academy of
Sciences, India (NASI).

\bibliographystyle{apsrev}
\bibliography{refs}
\end{document}